\documentclass[12pt]{iopart}

\usepackage{iopams}
\usepackage{setstack}
\usepackage{graphicx}
\usepackage{float}
\usepackage{bm}
\usepackage{verbatim}
\usepackage{tensor}
\usepackage{undertilde}
\usepackage{cite}
\usepackage{epstopdf}

\newcommand{\be}{\begin{equation}}
\newcommand{\ee}{\end{equation}}
\newcommand{\bea}{\begin{eqnarray}}
\newcommand{\eea}{\end{eqnarray}}

\DeclareMathAlphabet{\mathpzc}{OT1}{pzc}{m}{it}
\makeatletter
\DeclareRobustCommand{\text}{%
  \ifmmode\expandafter\text@\else\expandafter\mbox\fi}
\let\nfss@text\text
\def\text@#1{{\mathchoice
  {\textdef@\displaystyle\f@size{#1}}%
  {\textdef@\textstyle\f@size{#1}}%
  {\textdef@\textstyle\sf@size{#1}}%
  {\textdef@\textstyle \ssf@size{#1}}%
  \check@mathfonts
  }%
}
\def\textdef@#1#2#3{\hbox{{%
                    \everymath{#1}%
                    \let\f@size#2\selectfont
                    #3}}}
\makeatother

\begin{document}


\title[Cloud of strings in f(R) gravity]{Cloud of strings in f(R) gravity}

\author{J. P. Morais Gra\c ca$^{1, 2}$, Iarley P. Lobo$^{2}$ and Ines G. Salako$^{3}$}

\address{$^{1}$ Departamento de F\'{i}sica, Universidade Federal do Paran\'a, Curitiba, PR, Brazil}

\address{$^{2}$ Departamento de F\'{i}sica, Universidade Federal da Para\'{i}ba, Caixa Postal 5008, CEP 58051-970, Jo\~{a}o Pessoa, PB, Brazil}

\address{$^{3}$ Institut de Math\'ematiques et de Sciences Physiques (IMSP)\\
 01 BP 613 Porto-Novo, B\'enin}

\eads{\mailto{jpmorais@gmail.com}, \mailto{iarley\_lobo@fisica.ufpb.br}, \mailto{inessalako@gmail.com}}

\begin{abstract}
We derive the solution for a spherically symmetric string cloud configuration in a $d$-dimensional spacetime in the framework of $f(R)$ theories of gravity. We also analyze some thermodynamic properties of the joint black hole - cloud of strings solution. For its Hawking temperature, we found that the dependence of the mass with the horizon is significantly different in both theories. For the interaction of a black hole with thermal radiation, we found that the shapes of the curves are similar, but shifted. Our analysis generalizes some known results in the literature. 
\end{abstract}


\maketitle




%
%
\section{Introduction}

A promising theory for the unification of all known forces of nature is based on the idea that all known fundamental particles are, in fact, vibration modes of one-dimensional string objects \cite{Becker:2007zj}. Such strings could have been stretched at the epoch known as inflation, where the universe has experienced a period of a fast accelerated expansion. This mechanism could give rise to a type of cosmic string, a large-scale object that in principle could pervade our observed universe \cite{HenryTye:2006uv}.

Large-scale cosmic strings are also predicted by the theory of quantized fields. Such objects would originate from some phase transition in the early universe \cite{Hindmarsh:1994re}. The basic idea is that causally disconnected regions of the spacetime can develop different vacuum expectation values, and the regions where they overlap must be the regions of localized energy. Such localized energy can exist for domain walls, cosmic strings, and monopoles, among other possibilities. 

In Ref.~\cite{Letelier:1979ej}, Letelier proposed a model for a cloud of strings, an aggregation of one-dimensional objects in a defined geometrical pattern. A cloud of strings is analogous to a pressureless perfect fluid. However, due to the one-dimensional character of the string, its energy-momentum tensor presents a spacial component, thus manifesting a non-null pressure. This behavior implies that such an object should give rise to astrophysical and cosmological implications. Since then, several papers have been devoted to the study of string clouds in the General Relativity context \cite{Richarte:2007bx,Yadav:2009zza,Ganguly:2014cqa,Bronnikov:2016dhz,Barbosa:2016lse} and for alternative theories of gravity \cite{Herscovich:2010vr,Ghosh:2014dqa,Ghosh:2014pga,Lee:2014dha,Mazharimousavi:2015sfo,Graca:2016cbd}.

Our goal in this paper is to further explore these string clouds in the framework of $f(R)$ theories of gravity. In this theory, the scalar curvature in Einstein-Hilbert action, $R$, is replaced by a generic function, $f(R)$ \cite{Sotiriou:2008rp,Nojiri:2010wj,Nojiri:2017ncd}. The freedom to choose, in principle, any functional form for $f(R)$ allows us to find exact solutions in what would be a rather complicated set of coupled 4th-order non-linear differential equations. This freedom can also be seen as a drawback, since the obtained solutions can impose constraints on the final form of $f(R)$. A similar analysis was performed in Ref.\cite{Man:2013sf} for the case of a global monopole in four dimensions (whose energy-momentum tensor coincides with the string cloud one in 4D). However, the authors do not consider the $f(R)$-correction of the black hole entropy. Therefore in this work, besides generalizing the solutions and thermodynamic description for arbitrary spacetime dimensions, we consider such a $f(R)$-contribution to the area law and its thermodynamic implications.

In this paper we will focus on gravitational considerations of the cloud of strings in $f(R)$ gravity in a spherically symmetric spacetime of arbitrary dimensions. In Section 2 we will present the string cloud model as proposed by Letelier, considering the strings within a spherical distribution. In Section 3 we will briefly present the $f(R)$ model of gravity in $d$-dimensions, along with the most straightforward method to obtain exact solutions. In Section 4 we will derive the metric solution for the string cloud using the tetrad formalism. In Section 5 we will perform the thermodynamic analysis of the solution. Finally, in Section 6, we will present our conclusions.

\section{The cloud of strings}

In cosmology, we consider perfect fluids to model gas and dust of particles. A cloud of strings is a kind of analogue model, but with one-dimensional objects, which are extended along some defined direction. Such distributions of strings can exist in several geometrical shapes, such as planar, axisymmetric, or spherical. In this paper we will consider the last option.  

A particle generates a world line in spacetime, and a $d$-dimensional object generates a $(d+1)$-dimensional worldsheet. In the case of a single string, we say it generates a two-dimensional worldsheet $\Sigma$, that we can parametrize by two parameters: $\lambda^0$ and $\lambda^1$. The action for such a string is proportional to the determinant of the worldsheet, which is also known as the Nambu-Goto action, given by 

\begin{equation}
S_{GN} = m \int_{\Sigma} \sqrt{-\gamma} d\lambda^0 d\lambda^1, 
\label{NGaction}
\end{equation}
where $m$ is positive and is related to the tension of the string, and $\gamma$ is the determinant of the induced metric 

\begin{equation}
\gamma_{ab} = g_{\mu\nu} \frac{\partial x^\mu}{\partial \lambda^a} \frac{\partial x^\nu}{\partial \lambda^b}.
\end{equation}

We can also write the Nambu-Goto action using a spacetime bi-vector $\Sigma^{\mu\nu}$, given by

\begin{equation}
\Sigma^{\mu\nu} = \epsilon^{ab} \frac{\partial x^\mu}{\partial \lambda^a} \frac{\partial x^\nu}{\partial \lambda^b},
\end{equation}
in such a way that the Nambu-Goto action can be rewritten as

\begin{equation}
S_{GN} = m \int_{\Sigma} \sqrt{-\frac{1}{2} \Sigma_{\mu\nu} \Sigma^{\mu\nu}} d\lambda^0 d\lambda^1. 
\label{NGaction2}
\end{equation}

The energy-momentum tensor for one string can be calculated from the action (\ref{NGaction2}) in a straightforward way from the relation $T_{\mu\nu} = -2 \partial \mathcal{L} / \partial g^{\mu\nu}$. The energy-momentum tensor is then given by

\begin{equation}\label{Tmunu}
T^{\mu\nu} = m \frac{\Sigma^{\mu\sigma} \Sigma\indices{_\sigma^\nu}}{\sqrt{-\gamma}},
\end{equation}
and we can move from a single string to a string cloud by multiplying the energy-momentum tensor by a density, such as $T_{cloud}^{\mu\nu} = \rho T^{\mu\nu}$, where $\rho$ is the number density of the string cloud.

We will now consider a static, spherically symmetric string configuration. In this case, the only non-null component of the bi-vector is $\Sigma^{tr} = - \Sigma^{rt}$. Using Eq.~(\ref{Tmunu}), it can be shown that the only non-null components of the energy-momentum tensor of a string cloud are given by

\begin{equation}
T\indices{^t_t} = T\indices{^r_r} = m\rho\Sigma^{tr}\Sigma_{rt}/(-\gamma)^{1/2}.
\label{Trelation}
\end{equation}

We can now use the conservation law $\nabla_\mu T^{\mu\nu}=0$ to calculate the radial dependence of the components of the energy-momentum tensor. To do so, we must first define our metric. A $d$-dimensional static spherically symmetric metric is given by

\begin{equation}
ds^2 = -A(r) dt^2 + B(r) dr^2 + r^2 \gamma_{ij} dx^i dx^j,
\end{equation}
where $\gamma_{ij}$ is a $(d-2)$-dimensional sphere, and the indices $(i,j,...)$ run from $2$ to $d$. Following Ref.\cite{Letelier:1979ej}, in this metric the conservation of energy-momentum tensor leads to a ``conservation'' equation for $\Sigma^{\mu\nu}$ as $\partial_{\mu}[m\rho(-g)^{1/2}\Sigma^{\mu\nu}]=0$, whose solution, together with the expression for $\sqrt{-\gamma}$, which can be read from the identity $\Sigma^{\mu\alpha}\Sigma_{\alpha\beta}\Sigma^{\beta\nu}=\gamma\Sigma^{\nu\mu}$, leads to the non-null components of energy-momentum tensor with mixed indices

\begin{equation}
T^{t}_{\ t} =T^r_{\ r} = - \frac{\eta^2}{r^{d-2}},
\end{equation}
where we used Eq.~(\ref{Trelation}), and $\eta^2$ is a constant related to the total energy of the cloud of strings.
\par
Because of its radial dependence, there is an ``intra-string", but no ``inter-string" pressure, thus preventing the system from leaving the static configuration, which will be an important property for the stability of the black hole solution that we will find in the next section. The behavior of a string gas in a static configuration in a thermal context is discussed in Ref.\cite{Mertens:2014dia}.


\section{f(R) gravity}

In this section, we will briefly introduce the framework of $f(R)$ theories of gravity. For a broader review of the subject, see Refs.~\cite{Sotiriou:2008rp,Nojiri:2010wj,Nojiri:2017ncd}. The main idea of $f(R)$ gravity is to replace the Einstein-Hilbert action by a more general action. In $d$-dimensions it reads

\begin{equation}
S = \frac{1}{2 \kappa} \int d^dx \sqrt{-g} f(R) + S_m\, ,
\end{equation} 
where $S_m$ is the action for the matter fields, and $f(R)$ is an unknown function subject to the following constraints: $df(R)/dR > 0$ and $ d^2f(R)/dR^2 > 0$. The field equations can be derived from the above action. Varying it with respect to the metric, we obtain the field equations \cite{Gunther:2004ht}

\begin{equation}
F(R) R_{\mu\nu} - \frac{1}{2}f(R) g_{\mu\nu} - \nabla_{\mu} \nabla_{\nu} F(R) + g_{\mu\nu} \Box F(R) = \kappa^2 T_{\mu\nu},
\label{eom}
\end{equation}
where $F(R) = df(R)/dR$ and $T_{\mu\nu}$ is the energy-momentum tensor. We can clearly see that this set of equations are of fourth order in the metric, since the scalar curvature already contains two derivatives on the metric. The process of finding solutions for this set of equations, even for a simple function such as $f(R) = R + \alpha R^2$, is highly non-trivial and in general cannot be done analytically for a non-constant scalar curvature.

One method to solve Eq.~(\ref{eom}) is to apply what is known as the reconstructive approach. This idea originates from cosmological studies, where the aim was to reconstruct the $f(R)$ functional form based on the desired results for the scale factor of a FLRW metric. Within this approach, the $f(R)$ is treated as an independent field, and its functional form will be constrained \textit{a posteriori}.

Taking the trace of Eq.~(\ref{eom}), we obtain

\begin{equation}
F(R)\, R-\frac{d}{2}\, f(R)+(d-1)\, \Box F(R)=\kappa^2 T\, ,
\end{equation}
(where $T=g^{\mu\nu}T_{\mu\nu}$) and, substituting this expression in Eq.~(\ref{eom}), we derive

\begin{eqnarray}
F(R)R_{\mu\nu} - \nabla_\mu \nabla_\nu F(R) - \kappa^2 T_{\mu\nu} \nonumber\\
= \frac{1}{d}g_{\mu\nu}[F(R)R-\Box F(R) - \kappa^2 T].
\end{eqnarray} 

From the above equation, we can note that the combination below,

\begin{equation}
C_\mu = \frac{F(R)R_{\mu\mu} - \nabla_\mu \nabla_\mu F(R) - \kappa^2 T_{\mu\mu}}{g_{\mu\mu}},
\end{equation}
with fixed indices, is independent of the corresponding index, and the following relation, 

\begin{equation}
C_\mu - C_\nu = 0
\label{relation}
\end{equation}
gives us a new set of equations for any pair of $(\mu,\nu)$-indices. In the following section, we will use this formula to find an exact solution for the string cloud for an arbitrary spacetime dimension.


\section{Exact solution}

As mentioned in Section 2, a static spherically symmetric metric in $d$-dimensions is given by

\begin{equation}
ds^2 = -A(r) dt^2 + B(r) dr^2 + r^2 \gamma_{ij} dx^i dx^j,
\end{equation}
where $\gamma_{ij}$ is a $(d-2)$-dimensional sphere, and the latin indices $(i,j,...)$ run from $2$ to $d$. The approach we will use to deal with a submanifold of an arbitrary $(d-2)$ dimension is to define a set of orthonormal vielbeins, given by

\begin{eqnarray}
e^{(t)} = \sqrt{A(r)} dt, \hspace{10pt} e^{(r)} = \sqrt{B(r)} dr \hspace{10pt} \nonumber \\
\text{and} \hspace{10pt} e^{(a)} = e\indices{^{(a)}_i} dx^i = r e\indices{_0^{(a)}_{i}} dx^i,
\end{eqnarray} 
where $e\indices{_0^{(a)}}$ stands for the set of orthonormal vielbeins of the $(d-2)$-dimensional sphere. We can use the first Cartan structural equation, $de^{(A)} = - \omega\indices{^{(A)}_{(B)}} \wedge e^{(B)}$, to calculate the spin connection $\omega\indices{^{(a)}_{(b)}}$. We obtain

\begin{equation}
\omega\indices{^{(t)}_{(r)}} = \frac{1}{2} \frac{A'}{A \sqrt{B}} e^{(t)}, \hspace{10pt} \omega\indices{^{(a)}_{(r)}} = \frac{e\indices{_0^{(a)}}}{\sqrt{B}} \hspace{10pt} \text{and} \hspace{10pt} \omega\indices{^{(a)}_{(t)}} = 0,
\end{equation}
where $'$ means the derivative with respect to $r$, along with the fact that the vielbein $e\indices{_0^{(a)}}$ satisfies its own structural equation, given by $de\indices{_0^{(a)}} = - \omega\indices{^{(a)}_{(b)}} \wedge e^{(b)}$. 

The curvature $2$-form is defined as $\rho\indices{^{(A)}_{(B)}} = d \omega\indices{^{(A)}_{(C)}} + d \omega\indices{^{(C)}_{(B)}}$, and is given by

\begin{eqnarray}
&\rho\indices{^{(t)}_{(r)}} = \left(\frac{A''}{2 \sqrt{AB}} - \frac{A'^2 B}{4(AB)^{3/2}} - \frac{A'B'}{4AB}\right) dr \wedge dt\, ,\\
&\rho\indices{^{(t)}_{(a)}} = -\frac{A'}{2B \sqrt{A}} e_{0(a)i} dt \wedge dx^i\, ,\\ 
&\rho\indices{^{(r)}_{(a)}} = \frac{1}{2} \frac{B'}{B^{3/2}} e_{0(a)i} dr \wedge dx^i\, ,\\
&\rho\indices{^{(a)}_{(b)}} = \left(\rho\indices{_0^{(a)}_{(b)ij}} - \frac{e\indices{_0^{(a)}_{i}} e_{0(b)j} }{B}\right) dx^i \wedge dx^j,
\end{eqnarray}
where $\rho\indices{_0^{(a)}_{(b)ij}}$ is the curvature $2$-form of the $(d-2)$-dimensional sphere. Finally, we can calculate the Riemann tensor from the curvature using the vielbein, and contract its first and third indices to obtain the Ricci tensor, given by

\begin{eqnarray}
R\indices{^t_t} = - \frac{1}{4} \frac{-A'B'Ar + 2 A''ABr - A'^2Br + 2(d-2) A'BA }{A^2 B^2 r},
\\
R\indices{^r_r} = - \frac{1}{4} \frac{-A'^2 Br + 2A''BAr - A'B'Ar - 2(d-2) B'A^2}{A^2 B^2 r},
\\
R\indices{^i_i} = \frac{1}{2} \frac{-A'Br + B'Ar + 2(d-3)B^2A - 2(d-3)BA}{A B^2 r}\, .
\end{eqnarray}

We can now use Eq.~(\ref{relation}) to obtain our field equations for the functions $A(r)$ and $B(r)$. The $C_t - C_r$ and $C_\mu - C_t$ give us the following set of equations,

\begin{eqnarray}
-2 r F'' + rF' \beta + (d-2) F \beta = 0
\label{eqn1}
\end{eqnarray}
and
\begin{eqnarray}
(12 - 4d)AB - (12 - 4d)A + 4 \frac{F'}{F} rA - 2 \frac{F'}{F} A' r^2 - 2 A'' r^2  \nonumber \\
+ r^2 A' \beta + (6-2d) r A' - 2rA \frac{B'}{B} + \frac{4 \kappa \eta^2 AB}{F r^{d-4}} = 0\, ,
\label{eqn2}
\end{eqnarray}
where $\beta \equiv (A'/A + B'/B)$. 

To find exact solutions of the above equations, we must make some considerations. Following Ref.~\cite{Carames:2011uu}, we will consider the weak-field limit of the theory, i.e., that the metric functions are given by $A(r) = (1 + a(r))$ and $B(r) = (1 + b(r))$, with both $a(r) \ll 1$ and $b(r) \ll 1$.  We will also consider that our theory represents only a small deviation from General Relativity, $F(R(r)) = (1 + \psi(r))$, with $\psi(r) \ll 1$. With these considerations, Eqs.~(\ref{eqn1}) and (\ref{eqn2}) are given by

\begin{eqnarray}
- 2 r \psi'' + (d-2)(a' + b') = 0
\label{eqn3}
\end{eqnarray}
and
\begin{eqnarray}
(12 - 4d) b + 4r\psi' - 2 a'' r^2 + (6-2d) ra' \nonumber \\
- 2rb' + 4\kappa \eta^2 (1 + a + b - \psi) r^{4-d} = 0\, .
\label{eqn4}
\end{eqnarray}

With $b(r) = -a(r)$, Eq.~(\ref{eqn4}) can be exactly solved for $\psi = \psi_0 r$, and has as solution

\begin{eqnarray}\label{a-solution}
a(r) = \frac{D_1}{r^{d-3}} + D_2 r^2 - \frac{2\psi_0 r}{(d-2)} \nonumber\\
+ \kappa \eta^2 \left( \frac{\psi_0}{(d-3) r^{d-5}} - \frac{2}{(d-2) r^{d-4}}\right).
\end{eqnarray} 
\par
For $d=4$, we recover the solutions obtained in Ref.~\cite{Carames:2011uu} for the global monopole source. However, contrary to the global monopole case, the string cloud can exist and be defined for any spacetime dimension. In the absence of the strings, and taking $\psi \rightarrow 0$, we recover the Tangherlini-Schwarzschild metric with a cosmological constant. It is interesting to note that the correction due only to the $f(R)$ gravity is linear in the coordinate $r$ for any dimension, and its contribution decreases as the dimension increases due to the $\psi_0/(d-2)$ factor. This will be analyzed in the next section.

Let us stress that we did not rely on the functional form of the $f(R)$ function. To reconstruct it, we should calculate the scalar curvature as a function of the radial coordinate $r$, and invert it to find $r(R)$. Then, we integrate $F(R)$ on the variable $R$ to obtain $f(R)$. As mentioned earlier, in the reconstructive approach the $f(R)$ is obtained as a constraint of our previous choices.
\par
As mentioned before, the radial profile of the string cloud pressure is responsible the stability and staticity of our solution. Inside the horizon the picture changes and, due to the gravitational pull of the black hole towards the singularity, the staticity condition is violated, therefore the conditions inside the horizon should be treated carefully. However, since black hole thermodynamics is the subject of the next section, and since it does not rely on the inner structure of the solution, we leave a detailed treatment of this region for future investigations.



\section{Thermodynamics}
In this section we will study the thermodynamic properties of the black hole (BH) solution found from this configuration.\footnote{For a thermodynamic analysis of BHs in the brane world scenario, see Refs.~\cite{Chakraborty:2014xla,Chakraborty:2015bja}.} Using the solution (\ref{a-solution}), in the approximation $\kappa\eta^2\frac{\psi_0}{r^{d-5}}\ll 1$, we have
\begin{equation}\label{metric-function1}
A(r)=1+\frac{D_1}{r^{d-3}} + D_2 r^2 - \frac{2\psi_0 r}{(d-2)} -  \frac{2\kappa \eta^2}{(d-2) r^{d-4}}.
\end{equation}
For simplicity, we will discard the cosmological constant term by choosing $D_2=0$, and consider 
\be
D_1=-\frac{16\pi GM}{(d-2)\Omega_{d-2}}\, ,
\ee
in order to recover the Schwarzschild-Tangherlini metric \cite{Tangherlini:1963bw} in the appropriate limit, where
\begin{equation}
\Omega_{d-2}=\frac{2\pi^{(d-1)/2}}{\Gamma((d-1)/2)}
\end{equation}
is the area of the unit $(d-2)$-sphere \cite{Dennison:2010wd}.
\par
Since the horizon consists in the solution of $A(r_H)=0$, we find that the mass parameter of the BH as a function $r_H$ is
\begin{equation}\label{mass_rh}
M=\frac{(d-2)\Omega_{d-2}}{16\pi G}\left[r_H^{d-3}-\frac{2\psi_0}{d-2}r_H^{d-2}-\frac{2\kappa\eta^2}{d-2}r_H\right].
\end{equation}

\begin{figure}[H]
\includegraphics[scale=0.43]{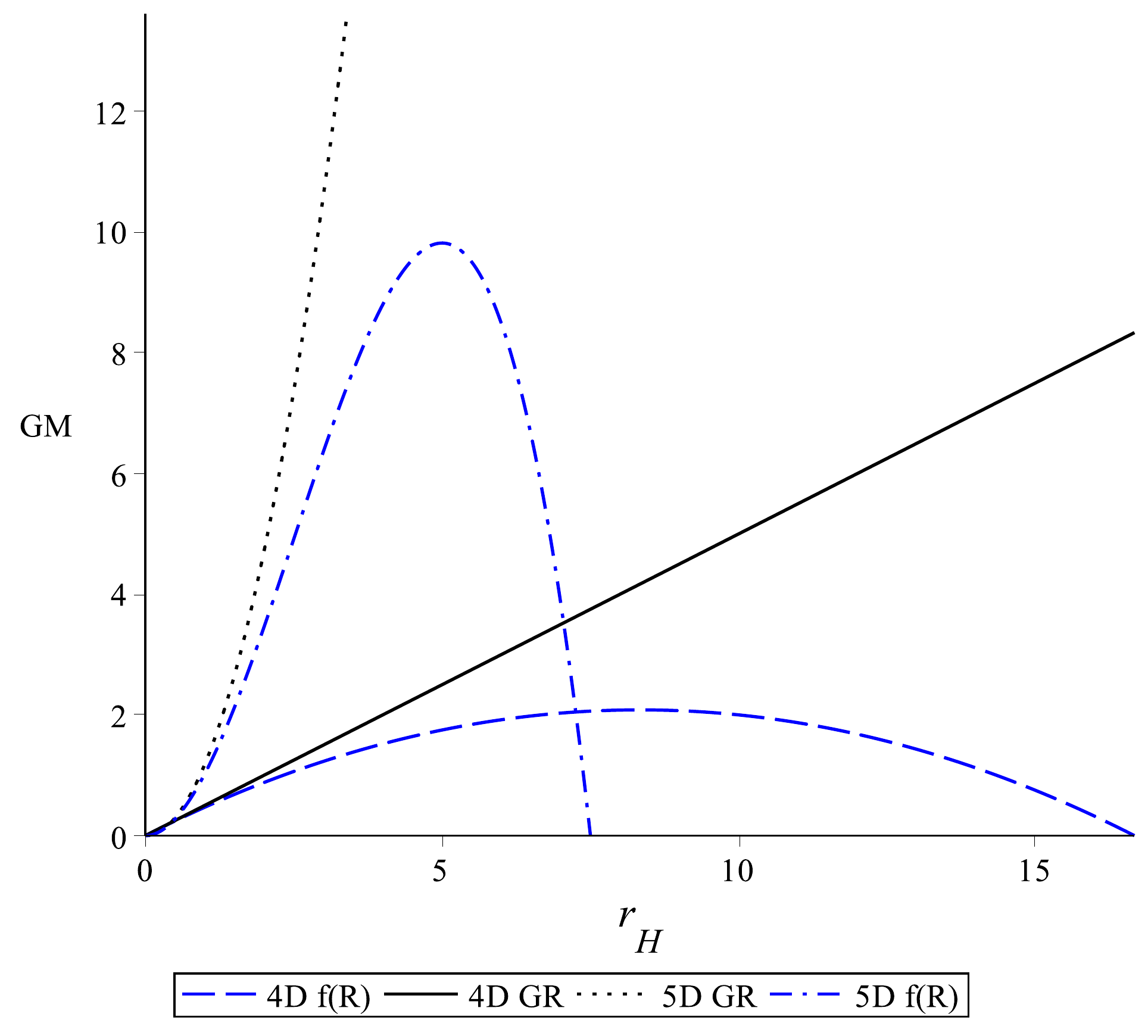}
\caption{The $4$-dimensional case for $f(R)$ (blue, solid line, where $\psi_0=0.06$) and General Relativity (dashed line); and the $5$-dimensional case for f(R) (blue, dash-dotted line, where $\psi_0=0.2$) and General Relativity (dotted line). We consider $ \kappa\eta^2=10^{-5}$.}
\label{MassxrH}
\end{figure}

In Fig.~\ref{MassxrH}, we show the behavior of the BH's mass parameter as a function of the horizon $r_H$. The shapes of the curves are considerably different, but there are some interesting common features, e.g., for General Relativity (GR), the energy is an ever growing function of the horizon, while for the $f(R)$ case, at some point the energy decreases with the growth of the horizon.

\begin{figure}[H]
\includegraphics[scale=0.43]{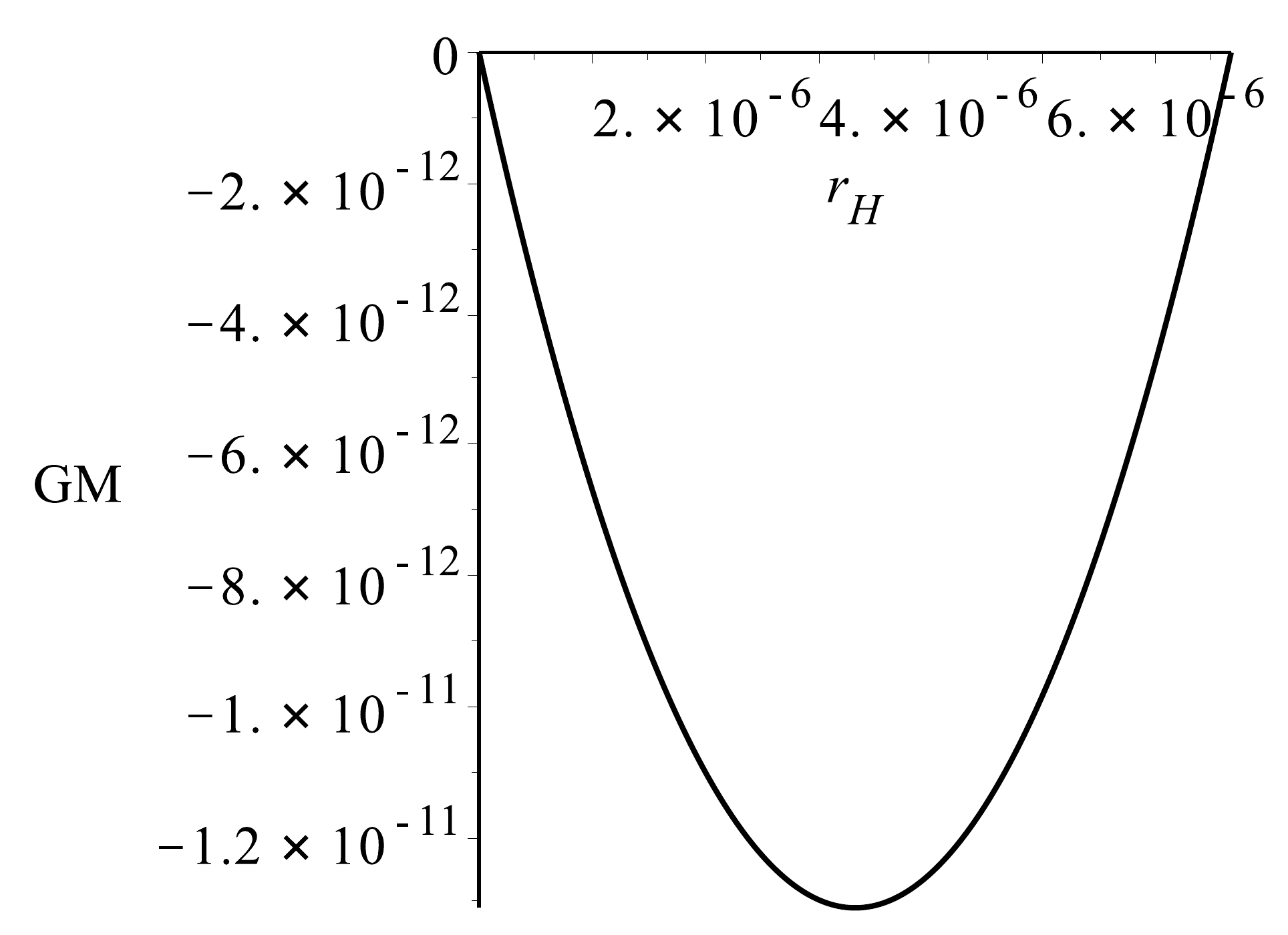}
\caption{The $5$-dimensional case, $\psi_0=0.2$ and $\kappa\eta^2=10^{-5}$. There is no significant difference between $f(R)$ and GR in this regime.}
\label{negativemass1}
\end{figure}

\begin{figure}[H]
\includegraphics[scale=0.43]{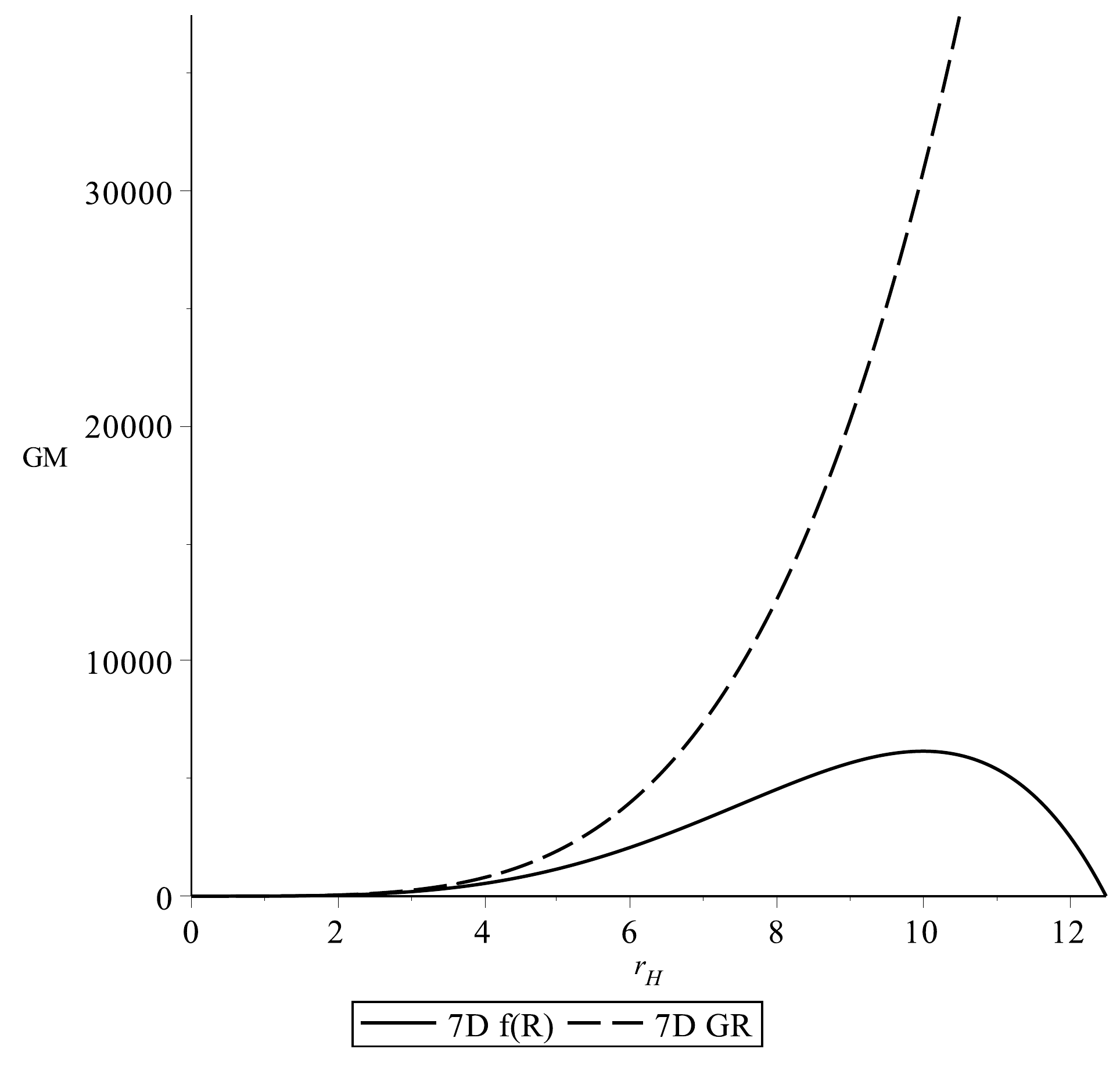}
\caption{The $7$-dimensional case for $f(R)$ (solid line, where $\psi_0=0.06$) and General Relativity (dashed line), with $ \kappa\eta^2=10^{-5}$.}
\label{MassxrH7D}
\end{figure}

\begin{figure}[H]
\includegraphics[scale=0.43]{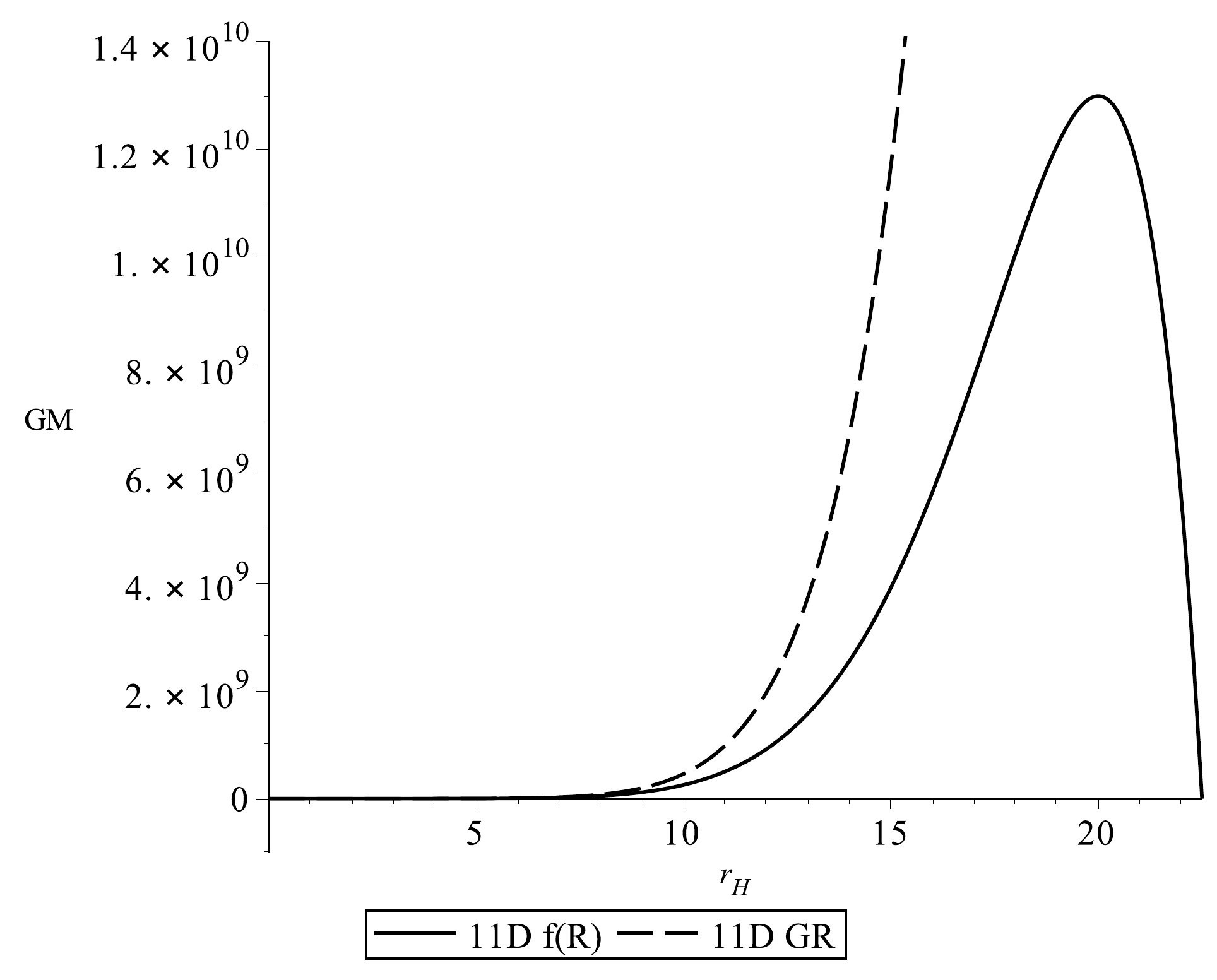}
\caption{The $11$-dimensional case for $f(R)$ (solid line, where $\psi_0=0.06$) and General Relativity (dashed line), with $ \kappa\eta^2=10^{-5}$.}
\label{MassxrH11D}
\end{figure}

Interestingly, the higher-dimensional case presents a region of negative mass parameter, as can be seen in Fig.~\ref{negativemass1}. This is due to the term $\kappa\eta^2 r_H$ in Eq. (\ref{mass_rh}), which, for $d>4$, becomes the dominant one for small values of $r_H$ (which is not the case for astrophysical observations), although some care should be taken since, for $d>5$, the limit that we are considering, $\kappa\eta^2\psi_0/r^{d-5}\ll 1$, may be violated for small values of $r_H$. For $d=4$, it has been shown that the analogue solution is not physical for $\psi_0 \gtrapprox 0.13$ \cite{Graca:2015jea}. It is possible that a similar behavior occurs for its higher dimensional counterparts. To see the behavior of the mass due to the change of dimensions, we plot the $7$- and $11$-dimensional cases in Figs.~\ref{MassxrH7D} and \ref{MassxrH11D}, respectively.

\par
According to Ref.~\cite{Emparan:2008eg}, the laws of black hole thermodynamics are valid in any number of dimensions. However, in $f(R)$, the entropy is proportional to the area of the event horizon corrected by a $f'(R)=F(R)$ factor (see Ref.~\cite{Faraoni:2010yi} and references therein):
\begin{equation}\label{entropy1}
S=\frac{k_B}{4G\hbar}\Sigma(r_H) F(R)|_{r=r_H}\, ,
\end{equation}
where $k_B$ is the Boltzmann constant, with dimensions of entropy.

The area of the BH is 
\begin{equation}
\Sigma(r_H)=\Omega_{d-2}r_H^{d-2}=\frac{2\pi^{(d-1)/2}}{\Gamma\left((d-1)/2\right)}r_H^{d-2}\, .
\end{equation}
It is worth stressing that our analysis is dimensionally coherent, since in $d$-dimensions $[\hbar G]=L^{d-2}$, i.e., it has dimensions of length to the power of $d-2$.

The Hawking temperature is a geometrical quantity given by
\begin{equation}
T_H(r_H)=\frac{\hbar}{k_B\Omega_{d-2}}\left[g^{tt}g^{rr}\frac{\partial g_{tt}}{\partial r}\right]_{r=r_H}.
\end{equation}
Using the mass parameter as a function of the horizon (\ref{mass_rh}), we find
\begin{eqnarray}\label{tinfty}
T_H(r_H)=\frac{\hbar}{k_B\Omega_{d-2}}\left[(d-3)r_H^{-1}- \frac{2\kappa\eta^2}{d-2}r_H^{3-d} - 2\psi_0\right]\, .\nonumber
\end{eqnarray}

For $\psi_0=0$, $\eta^2=0$ and $d=4$, we reproduce the temperature of the Schwarzschild BH ($r_H=2GM$)
\begin{equation}
T_{SC}=\frac{\hbar}{8\pi GMk_B}=\frac{1}{8\pi}\frac{E_P}{M}T_P\, ,
\end{equation}
where the Planck energy is $E_P=\sqrt{\hbar/G}$ and the Planck temperature is $T_P=E_P/k_B$.
\par
From these expressions it is possible to derive the dynamical mass which is the internal energy of this thermodynamic system. From the first law of BH thermodynamics, we have $dE=T_H\, dS$. Since we are expressing our quantities as functions of the horizon $r_H$, we can express this equation as
\begin{equation}
\frac{\partial E}{\partial r_H}=T_H\, \frac{\partial S}{\partial r_H},
\end{equation}
which furnishes, after an integration,
\begin{equation}
E(r_H)=\frac{1}{4G}\left[(d-2)r_H^{d-3}-2\kappa\eta^2r_H+\frac{(d^2-6d+7)}{d-2}\psi_0r_H^{d-2}\right].
\end{equation}


\subsection{Local Thermodynamics}
For a black hole in a thermal bath, we can calculate the local temperature $T_{loc}$. In fact, generalizing the prescription of Refs.~\cite{Tolman:1930zza,IBHP} for $d$-dimensions, we consider the presence of a perfect fluid that does not backreact on the BH solution
\be
T_{\mu\nu}=(\rho+p)v_{\mu}v_{\nu}+p\, g_{\mu\nu},
\ee
where $\rho$ is the energy density, $p$ is the pressure and $v^{\mu}$ is a unit, time-like vector, which is the $d$-velocity of a Killing observer.  We should stress that such an energy-momentum tensor is independent of our string cloud one (\ref{Tmunu}).  
From the conservation law $g^{\lambda\nu}\nabla_{\lambda}T_{\mu\nu}=0$, we have
\be\label{state1}
\partial_{\mu}p=-(\rho+p)\partial_{\mu}\ln (\alpha),
\ee
where $\alpha=\sqrt{-\xi^2}$, is the norm of a time-like Killing vector.
\par
For the analysis of the thermal bath, we consider the equation of state for radiation in a $d$-dimensional spacetime \cite{Alnes:2005ed,Amelino-Camelia:2016sru}:
\be
p=\frac{\rho}{d-1},
\ee
which implies that integrating Eq.~(\ref{state1}), we have
\be
\rho=\rho_{0}/\alpha^{d}.
\ee
Furthermore, according to the Stefan-Boltzmann law, the energy density of the thermal radiation in $d$-dimensions is given by \cite{Alnes:2005ed,Amelino-Camelia:2016sru}
\be
\rho \propto T_{loc}^d\, .
\ee
Hence, we must have $T_{loc} = T_{\infty}/\alpha$, for $\alpha=\sqrt{-\xi^2}=\sqrt{-g_{00}}$. Here, $T_{\infty}$ is the temperature measured at infinity: it is given by the Hawking temperature (\ref{tinfty}), $T_{\infty}=T_H$. The equation
\be
T_{loc}(r)=\frac{T_H}{\sqrt{-g_{00}(r)}}
\ee
determines Tolman's temperature $T_{loc}$. From this derivation we conclude that the expression of  Tolman's temperature is valid for any number of dimensions.
\par
In our case, $g_{00}=-A(r)$, thus
\begin{eqnarray}
&T_{loc}(r)=\frac{\hbar}{k_B\Omega_{d-2}}\left[(d-3)r_H^{-1}- \frac{2\kappa\eta^2}{d-2}r_H^{3-d} - 2\psi_0\right]&\nonumber\\
& \left[r\left(r^{d-4}-2\kappa\eta^2/(d-2)\right)-r_H\left(r_H^{d-4}-2\kappa\eta^2/(d-2)\right)\right.&\nonumber\\
&\left.+ \, 2\psi_0r_H^{d-2}/(d-2)-2\psi_0r^{d-2}/(d-2)\right]^{-1/2}\, \times r^{(d-3)/2}\, .&
\end{eqnarray}

\begin{figure}[H]
\includegraphics[scale=0.43]{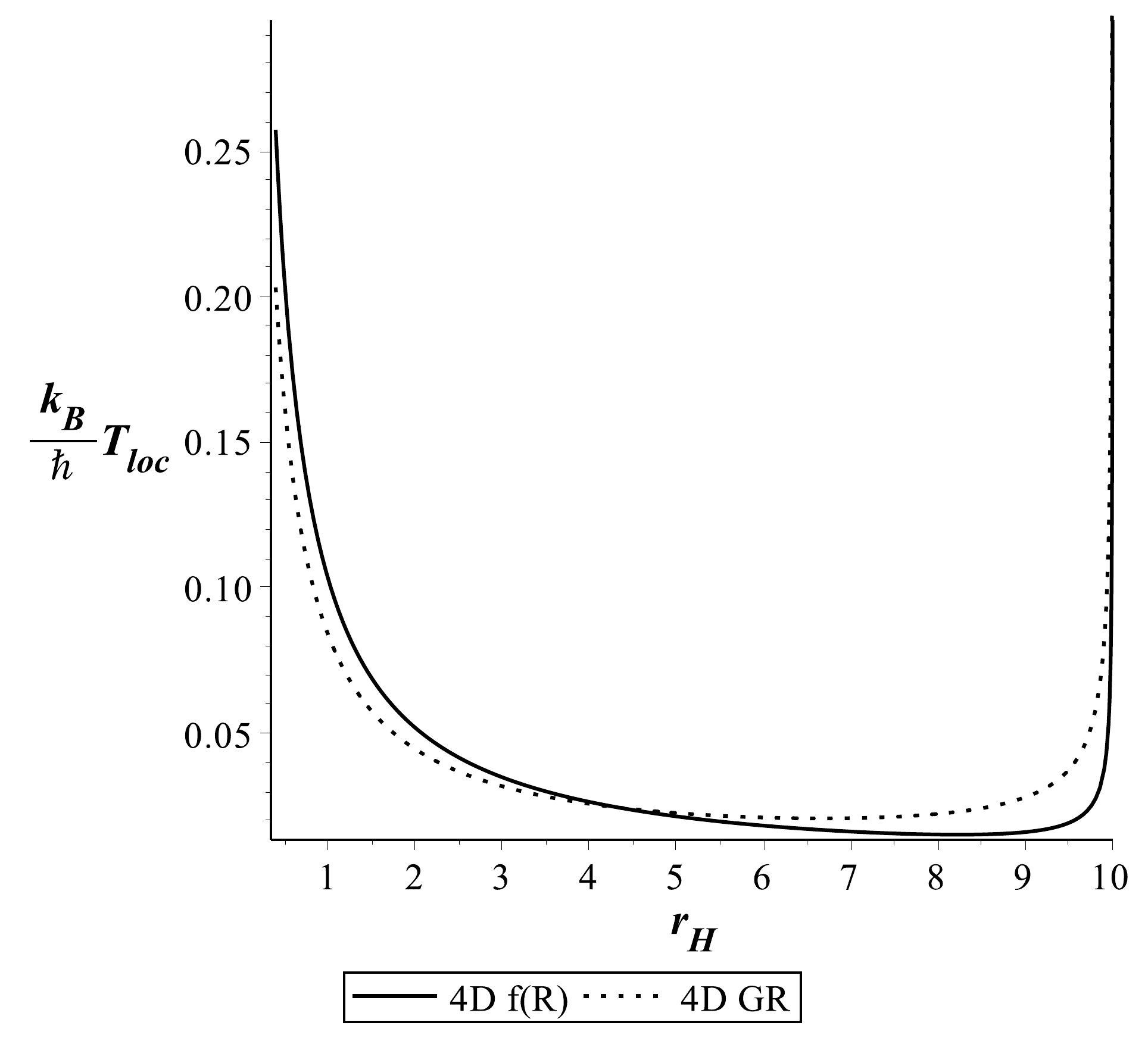}
\caption{The $4$-dimensional case, for General Relativity (dotted line) and $f(R)$ (solid line), with $\psi_0=0.04$, $\kappa\eta^2=10^{-5}$ and $r=10$.}
\label{4dtemp}
\end{figure}

\begin{figure}[H]
\includegraphics[scale=0.43]{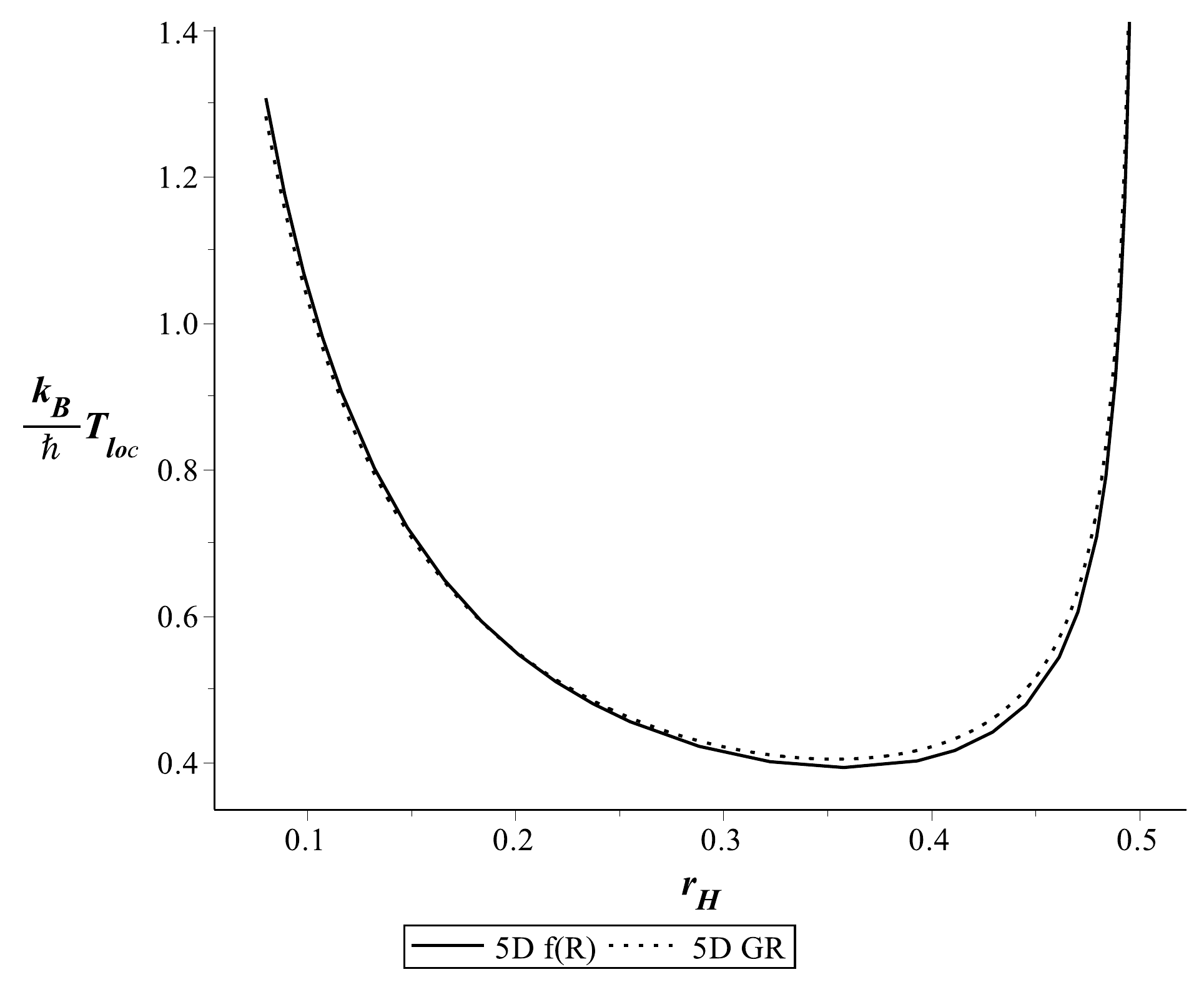}
\caption{The $5$-dimensional case, for General Relativity (dotted line) and $f(R)$ (solid line), with $\psi_0=0.2$, $\kappa\eta^2=10^{-5}$ and $r=1$.}
\label{5dtemp}
\end{figure}

\begin{figure}[H]
\includegraphics[scale=0.43]{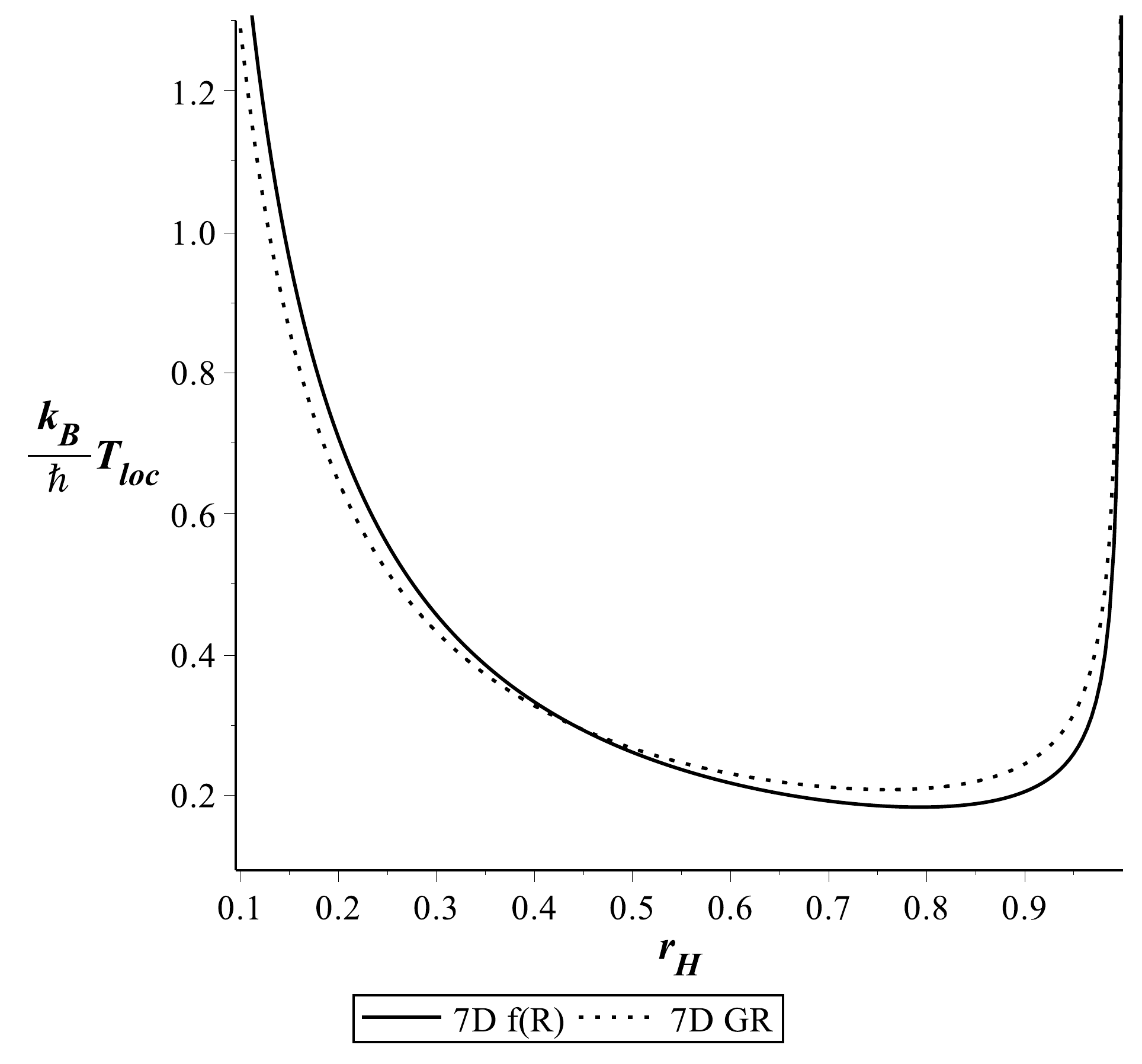}
\caption{The $7$-dimensional case, for General Relativity (dotted line) and $f(R)$ (solid line), with $\psi_0=0.7$, $\kappa\eta^2=10^{-5}$ and $r=1$.}
\label{7dtemp}
\end{figure}

\begin{figure}[H]
\includegraphics[scale=0.43]{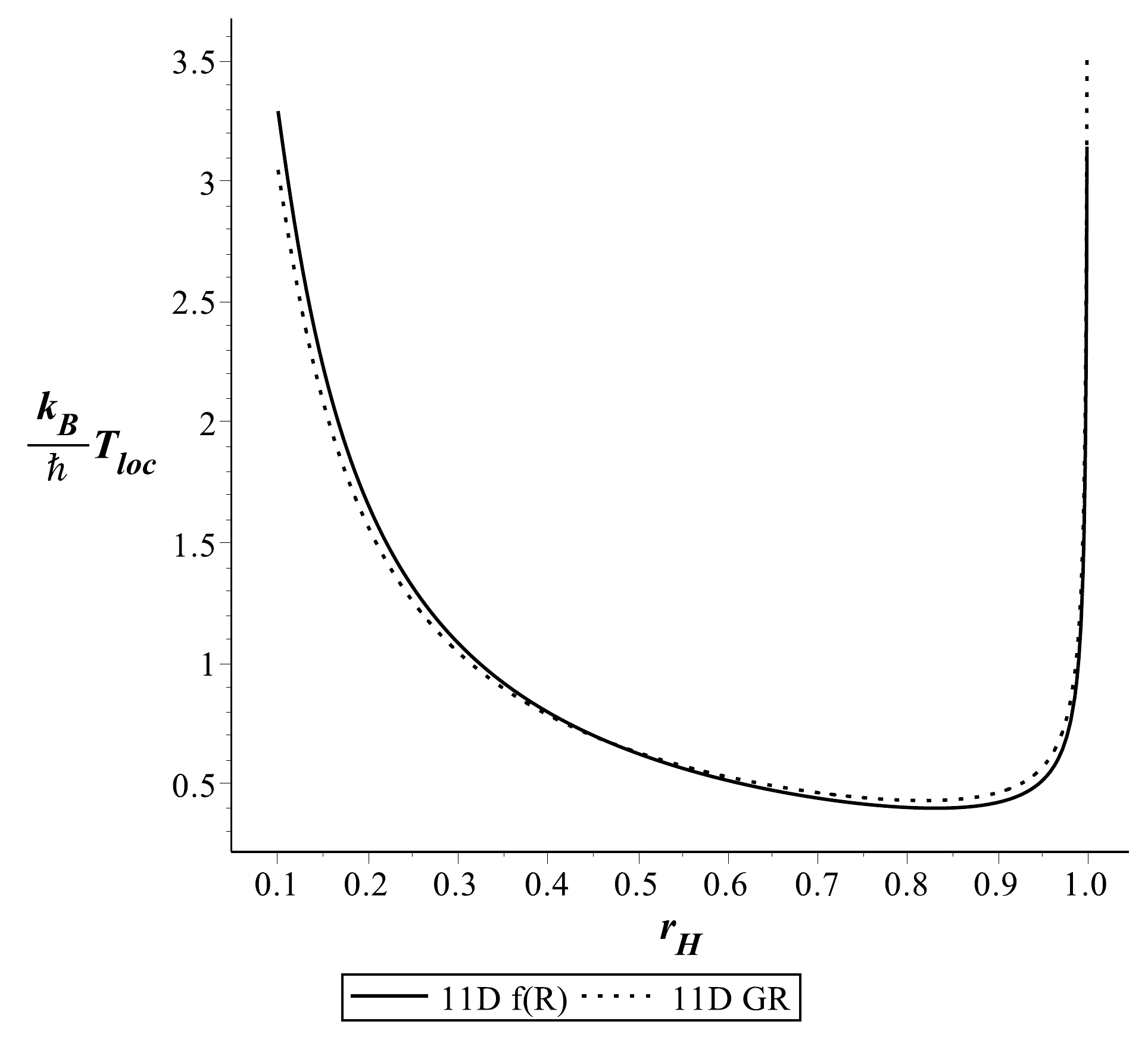}
\caption{The $11$-dimensional case, for General Relativity (dotted line) and $f(R)$ (solid line), with $\psi_0=0.8$, $\kappa\eta^2=10^{-5}$ and $r=1$.}
\label{11dtemp}
\end{figure}

The local temperature $T_{loc}$ as a function of the horizon $r_H$ in four dimensions is shown in Fig.~\ref{4dtemp}. There are minima in theses graphs, meaning that for a fixed distance from a BH, the minimum temperature allowed in $f(R)$ is smaller than that of GR. In Figs.~\ref{5dtemp}, \ref{7dtemp}, and \ref{11dtemp}, we consider the $5$-, $7$-, and $11$-dimensional cases respectively, and we verify that the graphs are similar to the previous case. In each case there is a minimum temperature, and the $f(R)$ black hole with a cloud of strings has a smaller minimum temperature than in GR.

From the first law of BH thermodynamics, we have
\be
dE_{loc}=T_{loc}dS,
\ee
where the entropy of the BH is given by Eq.~(\ref{entropy1}).
\par
Therefore, from $\partial E_{loc}/\partial r_H=T_{loc}\partial S/\partial r_H$, we derive
\begin{eqnarray}
&\frac{\partial E_{loc}}{\partial r_H}\Big |_r=\frac{1}{4G}\left[(d-2)(d-3)r_H^{d-4} - 2\kappa\eta^2 - 2\psi_0r_H^{d-3}(d^2-6d+7)\right]&\nonumber\\
&\left[r\left(r^{d-4}-2\kappa\eta^2/(d-2)\right)-r_H\left(r_H^{d-4}-2\kappa\eta^2/(d-2)\right)\right.&\nonumber\\
&\left.+ \, 2\psi_0r_H^{d-2}/(d-2)-2\psi_0r^{d-2}/(d-2)\right]^{-1/2}\times \, r^{(d-3)/2}\, .&
\end{eqnarray}
The heat capacity $C_{loc}$ can also be calculated as
\be
C_{loc}=\frac{\partial E_{loc}}{\partial r_H}\Big |_r\left(\frac{\partial T_{loc}}{\partial r_H}\Big |_r\right)^{-1}=\frac{H(r_H,r,\eta,d)}{W(r_H,r,\eta,d)},
\ee
where
\begin{eqnarray} 
\hspace*{-2.5cm} H(r,r_{H},\eta,d)=\\
\hspace*{-2.5cm} -2\left( dr_{H}\psi_{0}-r_{H}\psi_{0}+d-2 \right) k_{{B}} \left[ -1/4\, \left( d-2 \right)  \left( -2\,{r}^{d+1}\psi_{0}+ \left( d-2 \right) {r}^{d}-2\,{r}^{4}{\eta}^{2}\kappa \right)  \left( d-3 \right) {r_{H}}^{2\,d+1}\right.\nonumber\\
\hspace*{-2.5cm} -1/2\, \left( d-2 \right)  \left( 2\,{r}^{d+1}{\psi_{0}}^{2}-\psi_{0}\, \left( d-2 \right) {r}^{d}+{r}^{3}{\eta}^{2}\kappa\, \left( 2\,r\psi_{0}+d-2 \right)  \right) {r_{H}}^{2\,d+2}+1/4\,{r}^{3} \left( d-3 \right)  \left( d-2 \right) ^{2}{r_{H}}^{3\,d-2}\nonumber\\
\hspace*{-2.5cm}- \left( d-5/2 \right)  \left( d-2 \right) {r}^{3}\psi_{0}\,{r_{H}}^{3\,d-1}+{r}^{3}{\eta}^{2}\kappa\,\psi_{0}\, \left( d-1 \right) {r_{H}}^{2\,d+3}-{r_{H}}^{5+d}{r}^{d+1}{\eta}^{2}\kappa\,\psi_{0}\nonumber\\
\hspace*{-2.5cm}\left.+1/2\, \left(  \left( d-2 \right) {r}^{d}-2\,{r}^{4}{\eta}^{2}\kappa \right) {\eta}^{2}\kappa\,{r_{H}}^{5+d}+{r}^{3} \left( {\psi_{0}}^{2} \left( d-2 \right) {r_{H}}^{3\,d}+{r_{H}}^{d+6}{\eta}^{4}{\kappa}^{2} \right)  \right] \Omega_{{d-2}}\nonumber
\end{eqnarray}

and

\begin{eqnarray}
\hspace*{-2.5cm} W(r,r_H,\eta,d)=\\
\hspace*{-2.5cm} \hbar G \left[ -4\,{r}^{3}\psi\, \left( d-1 \right)  \left( d-2 \right)  \left( d-3 \right) r_{H}^{2\,d+1}+4\,{r}^{3}{\psi}^{2} \left( d-2 \right) ^{2}r_{H}^{2\,d+2}\right.\nonumber\\
\hspace*{-2.5cm}+4\, \left(  \left( d-2 \right) r_{H}^{d+3}-2\,r_{H}^{7}{\eta}^{2}\kappa \right) \psi\, \left( d-3 \right) {r}^{d+1} -2\, \left( d-2 \right)  \left(  \left( d-2 \right) {r}^{d}-2\,{r}^{4}{\eta}^{2}\kappa \right)  \left( d-3 \right) r_{H}^{d+3}\nonumber\\
\hspace*{-2.5cm}+16\, \left( d-5/2 \right) {r}^{3}{\eta}^{2}\kappa\,\psi\,r_{H}^{5+d} +{r}^{3} \left( d-1 \right)  \left( d-3 \right)  \left( d-2 \right) ^{2}r_{H}^{2\,d}-12\, \left( {r}^{3} \left( d-2 \right)  \left( d-3 \right) r_{H}^{4+d}\right.\nonumber\\
\left.\left.-1/3\, \left(  \left( {d}^{2}-5\,d+6 \right) {r}^{d}+2\,{r}^{3}{\eta}^{2}\kappa\, \left(  \left(r_{H}-r \right) d-5/2\,r_{H}+3\,r \right)  \right) r_{H}^{7} \right) {\eta}^{2}\kappa\right]\nonumber
\end{eqnarray}

For the case $d=4, \psi=0, \eta^2=0$, this reduces to the standard heat capacity for the Schwarzschild solution of GR
\begin{equation}
C_{loc}|_{Sch}=4\pi \frac{k_B}{\hbar G}\, \frac{r_H^2(r-r_H)}{3r_H-2r}
\end{equation}

\begin{figure}[H]
\includegraphics[scale=0.43]{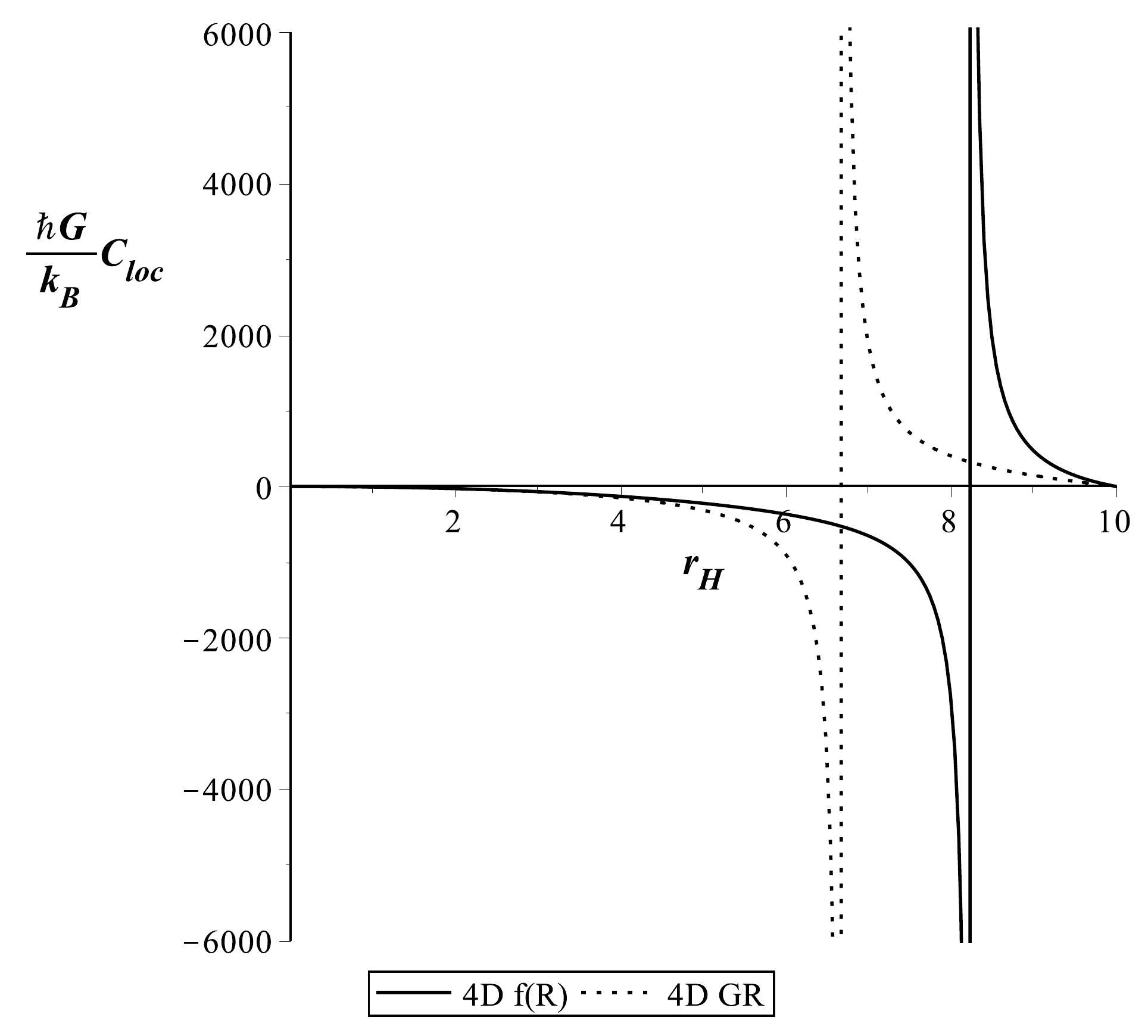}
\caption{The $4$-dimensional case, for General Relativity (dotted line) and $f(R)$ (solid line), with $\psi_0=0.04$, $\kappa\eta^2=10^{-5}$ and $r=10$.}
\label{4dCloc}
\end{figure}

\begin{figure}[H]
\includegraphics[scale=0.43]{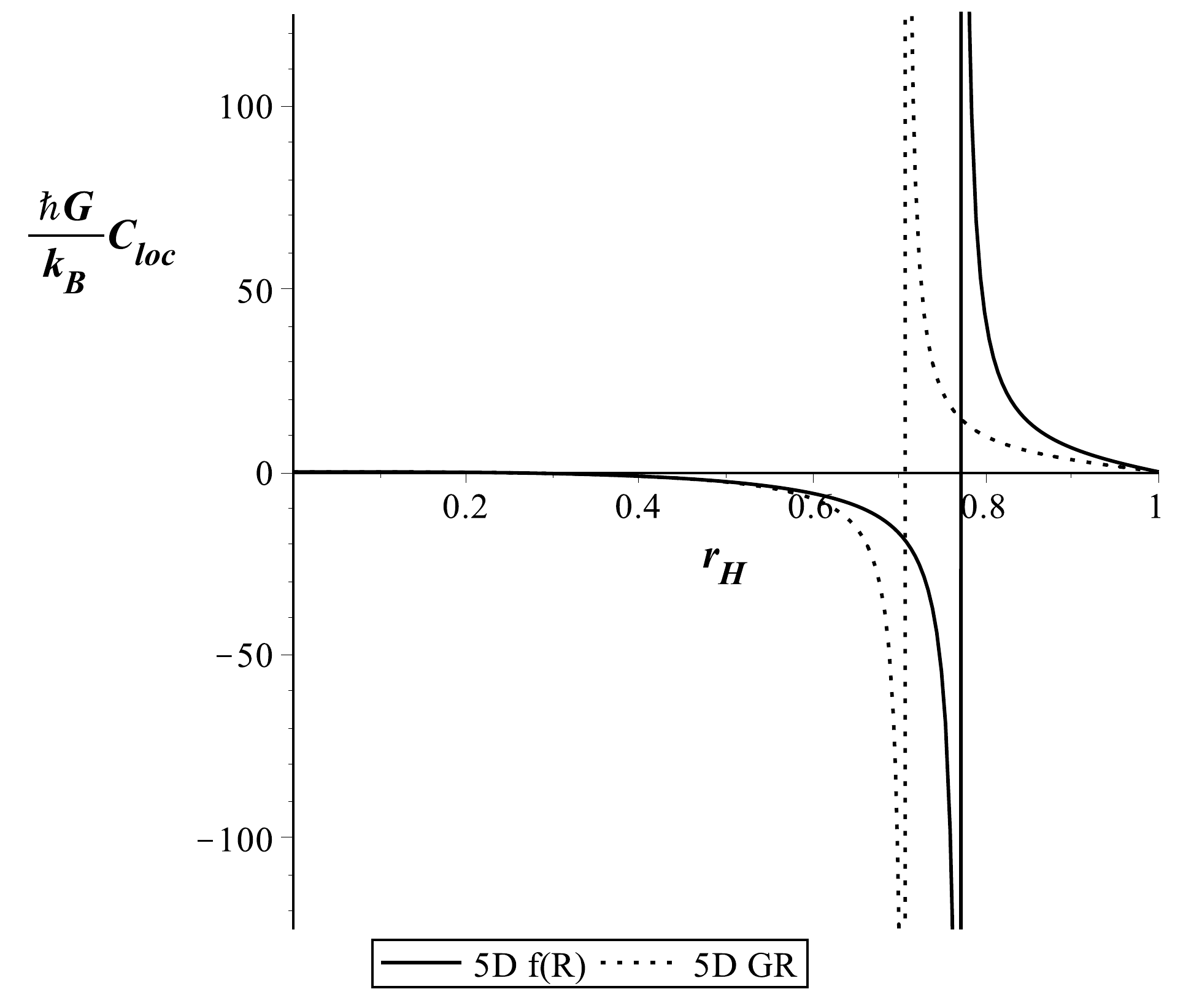}
\caption{The $5$-dimensional case, for General Relativity (dotted line) and $f(R)$ (solid line), with $\psi_0=0.5$, $\kappa\eta^2=10^{-5}$ and $r=1$.}
\label{5dCloc}
\end{figure}

\begin{figure}[H]
\includegraphics[scale=0.43]{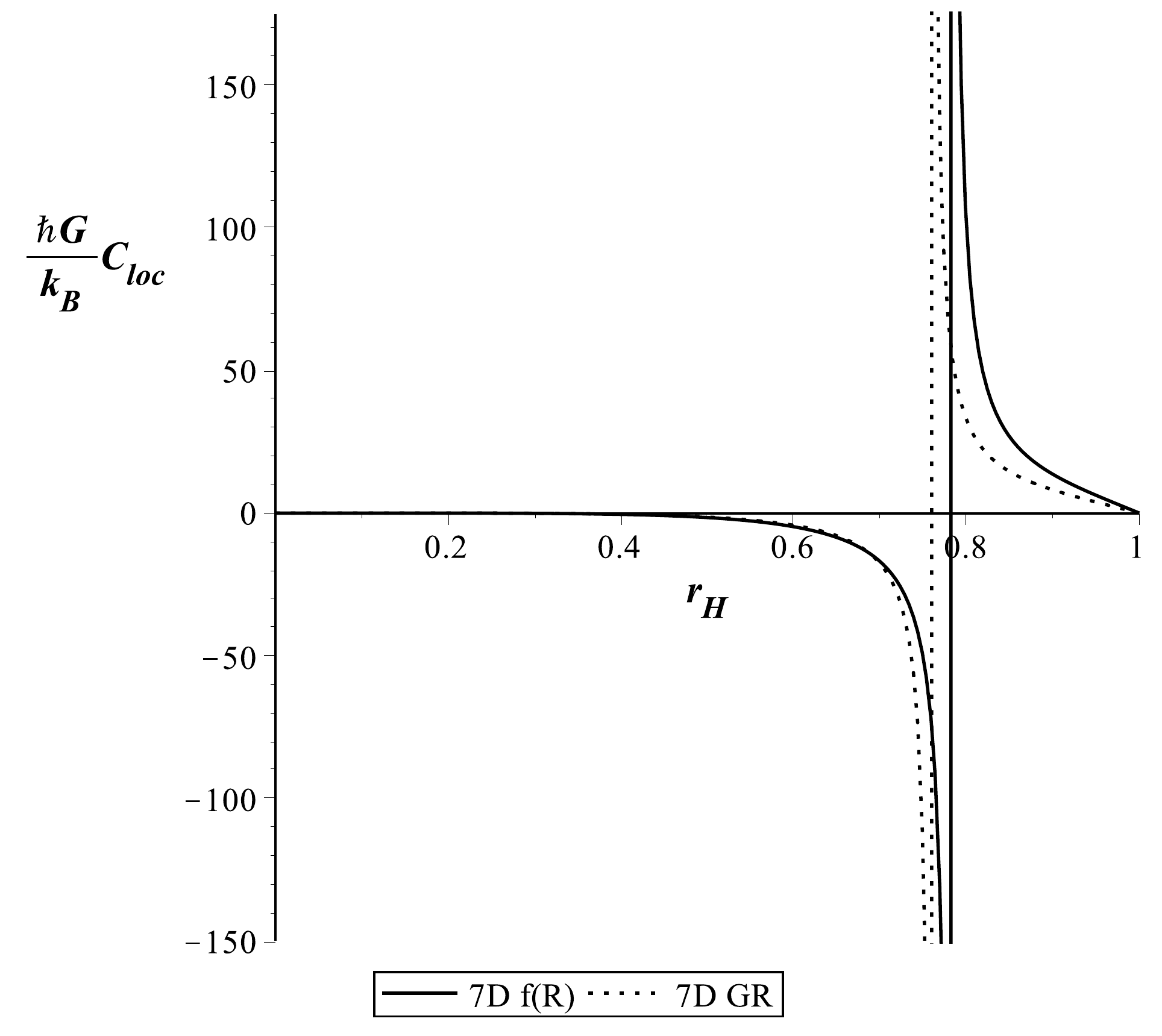}
\caption{The $7$-dimensional case, for General Relativity (dotted line) and $f(R)$ (solid line), with $\psi_0=0.5$, $\kappa\eta^2=10^{-5}$ and $r=1$.}
\label{7dCloc}
\end{figure}

\begin{figure}[H]
\includegraphics[scale=0.43]{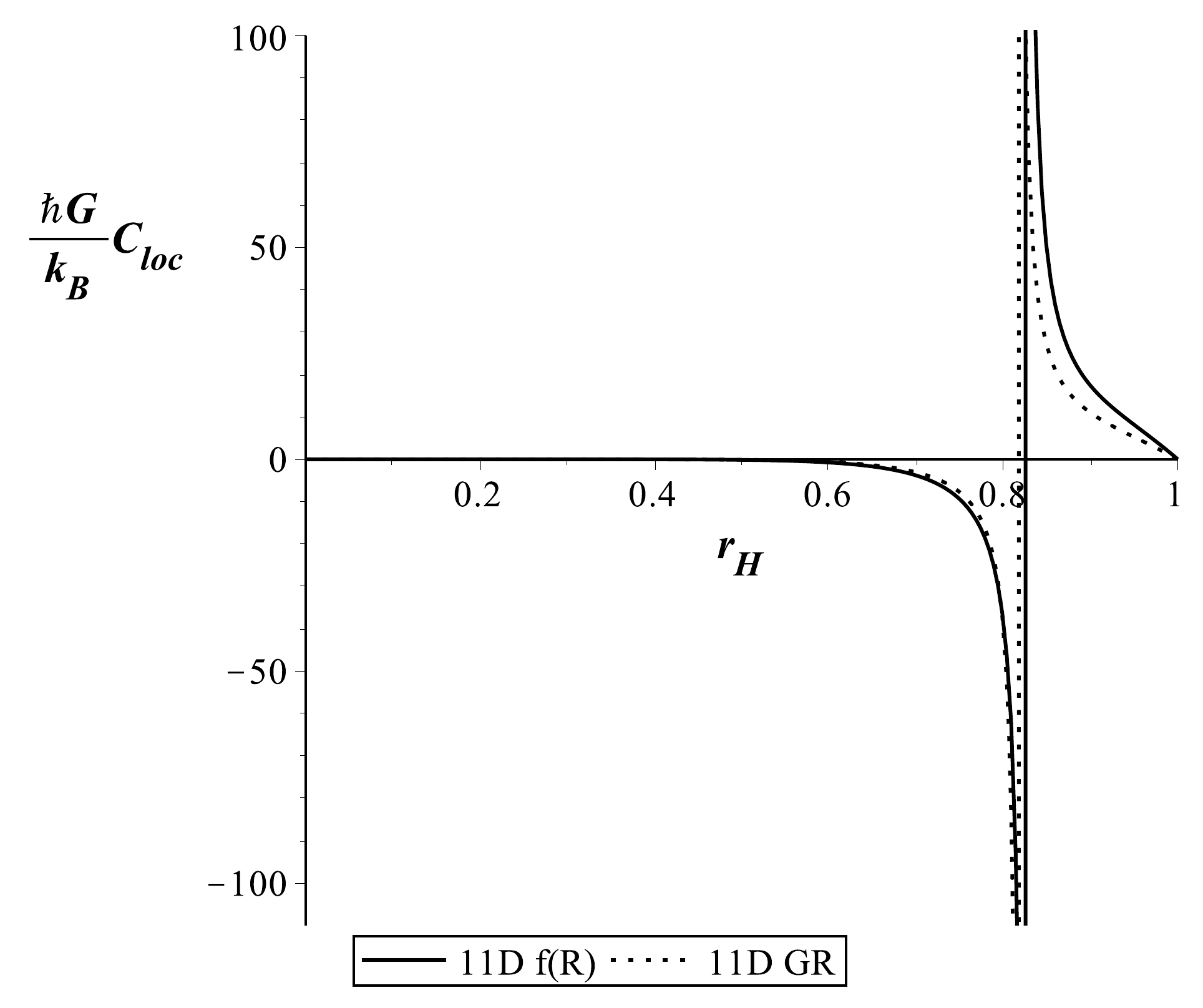}
\caption{The $11$-dimensional case, for General Relativity (dotted line) and $f(R)$ (solid line), with $\psi_0=0.5$, $\kappa\eta^2=10^{-5}$ and $r=1$.}
\label{11dCloc}
\end{figure}

We show the heat capacity for the $4$-, $5$-, $7$- and $11$-dimensional cases in Figs.~\ref{4dCloc}-\ref{11dCloc}, respectively. In all the cases,  stability is only possible for larger black holes due to negative heat capacities for small values of the horizon $r_H$. Their qualitative behavior is similar, independently of the number of dimensions. However, from Figs.~\ref{5dCloc}-\ref{11dCloc}, it can be seen that, as  expected from the shape of the metric (\ref{metric-function1}), the differences between $f(R)$ and GR reduce when the spacetime dimensions increase (for a fixed value of $\psi_0$).

\section{Concluding remarks}

In this paper we studied a string cloud configuration in the framework of $f(R)$ theories of gravity. Using the standard procedure of the reconstructive approach, we derived an exact solution for the $d$-dimensional spacetime metric generated by a symmetric configuration of such string cloud. For $d=4$, we recover the case of a global monopole, and in the limit with vanishing matter and where $f(R) \rightarrow R$, we recover the well-known Tangherlini-Schwarzschild metric. It is interesting to note that the metric correction due only to the $f(R)$ correction is linear in the radial coordinate, and decreases as we increase the spacetime dimensions.
\par
We also studied the thermodynamic properties of this solution, generalizing some previous results that were obtained in four dimensions and for a similar matter source \cite{Man:2013sf}. We calculated the horizon-dependence of the BH mass, demonstrating major differences between the GR and $f(R)$ cases; although within the same theory and independently of the number of dimensions, they share a similar qualitative behavior, for instance the unbounded mass-growth with the horizon $r_H$ for GR.
\par
We also considered the BH in a thermal bath, where we calculated the local temperature and the heat capacity. No significant qualitative difference between the theories was found (independently of the spacetime dimensions), except for a smaller temperature allowed in the $f(R)$ case (read from the local temperature analysis) and a shift in the heat capacity graphs for $f(R)$ in comparison to the GR case, which means that the minimum horizon radio (for a stable BH solution) is larger in $f(R)$ gravity than in GR.
\par
Our ansatz for the metric and the matter distribution of the string cloud obeys staticity and spherical symmetry, therefore we are imposing a time-independent configuration outside the horizon ``by hand". A non-static model with an accreting matter distribution, along with a dynamic string cloud atmosphere, would be a step towards a more realistic picture that might be used for observations. However, the existence of a solution of the field equations (following a similar path to those found in Refs.~\cite{Letelier:1979ej} and \cite{Kiselev:2002dx}), even in this restrictive model of a static configuration, instigates us to deepen this analysis. An interesting method would be the one followed by Ref.~\cite{Ganguly:2014cqa}.


\section*{Acknowledgements}
 IPL is supported by Conselho Nacional de Desenvolvimento Cient\'ifico e Tecnol\'ogico (CNPq-Brazil) by the grant No. 150384/2017-3. JPMG and IPL thank Coordena\c c\~ao de Aperfei\c coamento de Pessoal de N\'ivel Superior (CAPES) for financial support.

%
%

%
%
\end{document}